\documentclass[11pt,pdflatex]{amsart}

\usepackage{amsmath}
\usepackage[english,frenchb]{babel}
\usepackage{epsfig} 
\usepackage{graphicx}
\usepackage[dvips, bookmarks, colorlinks=true, plainpages = false, citecolor = red, urlcolor = blue, filecolor = blue]{hyperref}
 
\newtheorem{lem}{Lemma}

\newtheorem{theorem}{Theorem}

\begin{document}

\openup 1pt

\title[Farey Intervals]{Intervals Between Farey Fractions in the Limit of Infinite Level}
\author{Jan Fiala and Peter Kleban}
\address{Jan Fiala:  Department of Chemistry and Physics,
Lamar University,
Beaumont, Texas 77710.}
\email{jfiala@my.lamar.edu}
\address{Peter Kleban: LASST and Department of Physics \& Astronomy, University of Maine, Orono, ME 04469.}
\email{kleban@maine.edu}



\begin{abstract}
\begin{otherlanguage}{french}
\begin{center}
\large
\end{center}
La s\'{e}quence Farey modifi\'{e}e consiste, \`{a} chaque niveau $k$, des fractions rationelles  $r_k^{(n)}$, avec $n=1, 2, \dots,2^k+1$.  Nous consid\'{e}rons $I_k^{(e)}$, la longueur totale d'une partie des intervalles alternants entre les fractions  Farey qui paraissents pour la premi\`{e}re fois (soit sont ``nouveaux") \`{a} niveau $k$
$$I^{(e)}_k := \sum_{i=1}^{2^{k-2}} \left(r_k^{(4i)}- r_k^{(4i-2)}\right) \;.$$
Nous d\'{e}montrons que  $\liminf_{k\to \infty} I_k^{(e)}=0$, et posons la conjecture que $\lim_{k \to \infty}I_k^{(e)}=0$. Cette propri\'{e}t\'{e} g\'{e}ometrique simple des fractions Farey se montre assez subtile, et ne para\^{i}t pas avoir une interpr\'{e}tation  \'{e}vidente.  La conjecture \'{e}quivaut $\lim_{k \to \infty}S_{k}=0$, ou $S_{k}$  est la somme des carr\'{e}es inverses des d\'{e}nominateurs nouveaux \`{a} niveau $k$, $S_{k}:=\sum_{n=1}^{2^{k-1}} 1/\left(d_k^{(2n)} \right)^2$.   Notre preuve emploie des bornes pour la longueur des intervalles entre les fractions  Farey  en fonction des longueurs de leurs intervalles  ``parents" aux plus bas niveaux.   
\\
\end{otherlanguage}
\begin{otherlanguage}{english}
\begin{center}
A{\tiny BSTRACT}
\end{center}
The  modified Farey sequence consists, at each level $k$, of rational fractions $r_k^{(n)}$, with $n=1, 2, \dots,2^k+1$.  We consider $I_k^{(e)}$, the total length of (one set of) alternate intervals between  Farey fractions that are new (i.e., appear for the first time) at level $k$
$$I^{(e)}_k := \sum_{i=1}^{2^{k-2}} \left(r_k^{(4i)}- r_k^{(4i-2)}\right) \;.$$
We prove that  $\liminf_{k\to \infty} I_k^{(e)}=0$, and conjecture that in fact $\lim_{k \to \infty}I_k^{(e)}=0$. This simple geometrical property of the Farey fractions turns out to be surprisingly subtle, with no apparent simple interpretation.  The conjecture is  equivalent to $\lim_{k \to \infty}S_{k}=0$, where $S_{k}$  is the sum over the inverse squares of the new denominators at level $k$, $S_{k}:=\sum_{n=1}^{2^{k-1}} 1/\left(d_k^{(2n)} \right)^2$.   Our result makes use of bounds for  Farey fraction intervals in terms of their ``parent" intervals at lower levels.  
\end{otherlanguage}

\end{abstract}

\maketitle

\vspace{2 pc}

\section{Introduction}\label{treti}

 The Farey fractions (modified Farey sequence) may be defined as $r_k^{(n)}:=\frac{n_k^{(n)}}{d_k^{(n)}}$ , with $\text{gcd}(n_k^{(n)},d_k^{(n)})=1$, and
$n$ denoted the {\it order} of the Farey fraction at level $k$. Level $k=0$ consists
 of the two fractions $\left\{ \frac{0}{1},\frac{1}{1} \right \}$. 
 Succeeding levels are generated by keeping
all the fractions from level $k$ in level $k + 1$, and including new fractions.
  The new fractions at level $k+1$ are defined via
 $ d_{k+1}^{(2n)}:=d_k^{(n)}+d_k^{(n+1)}$ and
 $ n_{k+1}^{(2n)}:=n_k^{(n)}+n_k^{(n+1)}$,
 so that\\
$k=0 \quad  \left \{ \frac{0}{1},\frac{1}{1} \right \}$\\
$k=1 \quad \left \{ \frac{0}{1},\frac{1}{2},\frac{1}{1} \right \}$\\
$k=2 \quad \left \{ \frac{0}{1},\frac{1}{3},\frac{1}{2},\frac{2}{3},
\frac{1}{1} \right \}$, etc.\\[.1cm]
Note that $n=1,\ldots,2^k+1$. 
 
 It follows that the fractions at a given level are in increasing order.
 
The Farey fractions may also be defined using products of the $2 \times 2$ matrices  $A=\left({1\atop 1}{0\atop 1}\right)$ and $B=\left({1\atop 0}{1\atop 1}\right)$  (see \cite{KO} for details).

Our main result concerns the sum of lengths of  half of the intervals between ``new" Farey fractions.   Theorem \ref{thm1}  proves that  the lim inf of this sum vanishes in the limit of infinite level $k$. Based on numerical evidence, we also conjecture that the limit of this sum vanishes.  This very simple geometric property of the Farey fractions is not very apparent.  The intervals chosen are alternating, and there seems no obvious reason why the sum of their lengths should vanish in this limit.

In this paper, we focus on  the ``Farey tree", which 
means retaining only the $2^{k-1}$ even Farey fractions at each level $k>1$.  These are exactly the new fractions at each level.  In our notation they are of even order, i.e., $r_k^{(2n)}$
so for each level $k>1$ we obtain the set $$\{r_k^{(2n)}|\,n=1,\ldots,2^{k-1} \}\;.$$  

\begin{figure}[h]
\centering\includegraphics[width=8.0cm]{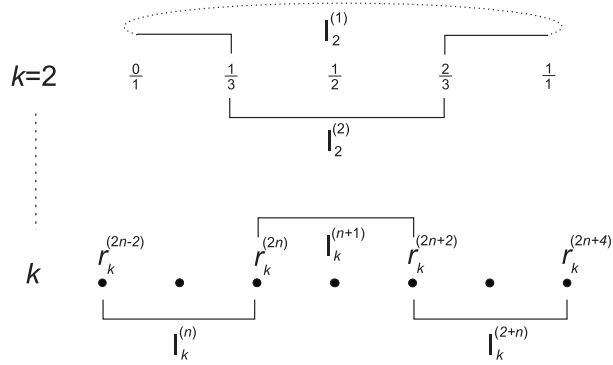}
\caption{Definition of Farey fraction intervals.  Even (odd) intervals are shown via a line below (above) the fractions (note that in the lower diagram, $n$ is even).  The dashed line in the upper diagram indicates the combining of the two ``extra" intervals into $I_k^{(1)}$.  In this figure and those below the intervals are not to scale.}\label{FFI}
\end{figure}

The lengths of the intervals between \emph{even} (new) Farey fractions at every level $k > 1$ are denoted
\begin{eqnarray}\label{Int}
I^{(n)}_k:= \left( r_k^{(2n)}- r_k^{(2n-2)} \right) > 0 \;,
\end{eqnarray}
where $n=2, 3, 4,\ldots, 2^{k-1}$.  In what follows, for brevity, we abuse the terminology slightly and refer to $I^{(n)}_k$ as an interval. When $n$ itself is even (i.e., $n=2, 4,\ldots, 2^{k-1}$), we refer to these as \emph{even intervals}.  (Note that there are $2^{k-2}$ even intervals at level $k$.) The complementary intervals in the unit interval $[0,1]$ i.e., those with  $n$ odd ($n=3, 5,\ldots, 2^{k-1}-1$), including the two extra intervals at the ends of the unit interval, namely $\left( r_k^{(2)}- r_k^{(1)} \right)$ and $\left( r_k^{(2^{k}+1)}- r_k^{(2^{k})} \right)$, are the \emph{odd intervals}. (Note that odd and even intervals alternate.) From the definition of the Farey fractions it is easy to verify that each of the extra intervals has length $1/(k+1)$. Thus we combine them and define 
\begin{equation}\label{Int1}
I_k^{(1)}:=2 /(k+1) \;,
\end{equation}
see Fig. \ref{FFI}.  (As a result there are  $2^{k-2}$ odd intervals at level $k$, the same as the number of even intervals.)

Next we define total length of even intervals at level $k$ 
\begin{eqnarray}\label{EvenInt}
I^{(e)}_k := \sum_{i=1}^{2^{k-2}}I^{(2i)}_k \;,
\end{eqnarray}
and similarly for the odd intervals  
\begin{eqnarray}\label{OddInt}
I^{(o)}_k:=\sum_{i=1}^{2^{k-2}}I^{(2i-1)}_k\;.
\end{eqnarray}

It follows that
\begin{eqnarray}\label{InCon}
I^{(e)}_k+I^{(o)}_k=1 \;,
\end{eqnarray}
for all $k\ge2$.

The quantity
\begin{equation}\label{Sdef}
S_{k}:=\sum_{n=1}^{2^{k-1}} \frac1{\left(d_k^{(2n)} \right)^2} \;,
\end{equation}
is the sum over the inverse squares of the new denominators at level $k$.  As we will see, $S_{k}$ is closely related to $I^{(e)}_k$  .

In the next section, by identifying intervals at different levels and examining their evolution from level to level,  we prove our main result
\begin{theorem}\label{thm1} 
\begin{equation}\label{liminfI}
\liminf_{k \to \infty} I_k^{(e)}=0 \;.
\end{equation}
\end{theorem}

Numerical evidence (see the last paragraph in Section \ref{FT}) then leads us to the \\ \\
{\bf Conjecture.}
\begin{equation} \label{limI} 
\lim_{k \to \infty} I_k^{(e)}=0 \;.
\end{equation}

Section \ref{relP} contains some further remarks concerning this conjecture.

\section{Proof of Theorem \ref{thm1}}\label{FT}
In this section we  prove   Theorem \ref{thm1} i.e.,  $\liminf_{k \to \infty} I_k^{(e)}=0$.  The key step is  Lemma \ref{lem7}, which  bounds an arbitrary odd interval  in terms of its ``parent" even interval at a lower level. In addition, at the end of the section we present numerical evidence for the  Conjecture (\ref{limI}).

It is convenient to use the full set of Farey fractions, even though only the even ones enter $I_k^{(e)}$ (see (\ref{EvenInt})). As mentioned, including $\frac01$ and $\frac11$, there are $2^k +1$ fractions at level $k \ge 1$.  In our notation, at a given level $k \ge 1$, the even-numbered fractions are new, having been ``born" at that level, while the odd-numbered ones are kept from the preceding level. 
Recall, also, that  the intervals in (\ref{EvenInt}) are exactly  the $I_k^{(n)}$ with $n$ even.  

\begin{lem} \label{lem2}
For any $k > 2$
we have the bounds  
\begin{eqnarray}\label{OddIntB}
\frac{2}{k+1}+\frac{2}{3}\sum_{j=1}^{k-2}I^{(e)}_{k-j}\frac{2j+3}{(j+1)(j+2)}\le I^{(o)}_k < 1\;,
\end{eqnarray}
and
\begin{eqnarray}\label{OddIntUpB}
\frac{2}{k+1}+\sum_{j=1}^{k-2}I^{(e)}_{k-j}\frac3{2j+3}\ge I^{(o)}_k \;.
\end{eqnarray}
\end{lem}
The rightmost inequality in  (\ref{OddIntB}) follows immediately from (\ref{InCon}).

 Now clearly $I^{(e)}_k > 0$ for any finite value of $k$.  On the other hand, if there were an $\epsilon > 0$ such that  $I^{(e)}_k \ge \epsilon$ for all $k$, the left hand side of (\ref{OddIntB}) would diverge as $k \to \infty$. Thus  Lemma \ref{lem2} implies  Theorem \ref{thm1}.

\begin{figure}[h]
\centering\includegraphics[width=10.0cm]{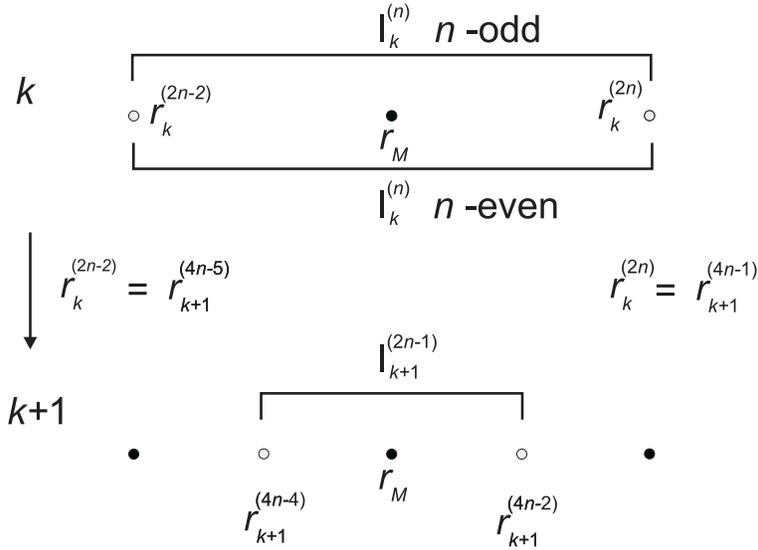}
\caption{Interval changes between levels of Farey fractions. The new Farey fractions (i.e., those ``born" at level $k+1$) are depicted by empty circles.  Note that any interval, even \emph{or} odd, at level $k$ gives rise to an odd interval at level $k+1$.}\label{FFL.eps}
\end{figure}

We prove  Lemma \ref{lem2}  by understanding how even and odd intervals evolve from level $k$ to level $k+1$ (see Fig. \ref{FFL}).  
To proceed, first note that the $n$th Farey fraction at level $k$ will be at order $2n-1$ in 
level $k+1$. Therefore we have
\begin{lem} \label{lem3}
{\it The transformation of the order for any Farey fraction in going from level 
$k\to k+l$ is}
\begin{eqnarray}\label{IndexTrans}
n\to 2^l(n-1)+1\;.
\end{eqnarray}
\end{lem} \qed

The first step in our argument involves identifying intervals at successive levels. We do this via their ``middle" Farey fractions.  In a slight abuse of notation, let $r_M$ be the ``middle" Farey fraction  of the interval $I^{(n)}_k$ at level $k$, i.e., $r_M = r_k^{(2n-1)}$  (see (\ref{Int})). We use this to identify any set of intervals at different levels with the same ``middle" fraction $r_M$.  Thus, since $r_M$ is necessarily of odd order, an interval at level $k$
is the  ``parent"  of the odd interval at level $k+1$ with the same $r_M$.

 It follows that any even interval  $I_k^{(2m)}$  at level $k$ will produce a (necessarily smaller) odd interval at level $k+1$, with $r_M=r_k^{(4m-1)} = r_{k+1}^{(8m-3)}$.   Similarly, any odd interval produces
an (odd) interval with the same  $r_M$ at the next level (see Fig. \ref{FFL}). In addition, there are new even intervals that are born at each level.  Their ``middle" fractions   are the ends of the even intervals from the previous level. 

At level $k=2$, we have one even interval, which lies between the two ``extra" intervals comprising $I_2^{(1)}$ at the ends of the unit interval. It follows that \emph{all} odd intervals at level $k > 2$ (except $I_k^{(1)}$) are born from even intervals
at some previous level. Further, every odd interval shrinks from level to level while preserving its ``middle" Farey fraction.   

This establishes
\begin{lem} \label{lem3}
For any level $k > 2$,  the unit interval is covered by a set of $2^{k-1}-1$ alternating even and odd intervals plus the two ``end" intervals comprising $I_k^{(1)}$.  The even intervals are ``newborn", while each of the odd intervals (except $I_k^{(1)}$) is the offspring of an even interval born at a previous level. 
\end{lem}

The next step is to determine what fraction of a given interval at level $k$ remains at level $k+1$. In doing this, it is useful to recall that the difference between any two successive fractions at a given level is  $r_{k}^{(n+1)}-r_{k}^{(n)}=\frac1{d_{k}^{(n+1)}d_{k}^{(n)}}$ (see for example \cite{KO,FKO}). 

Now consider an arbitrary even interval $I_{k+2}^{(2n)}$ at level $k+2$ (for $k \ge 0$; note that $n=1, 2, \ldots, 2^k$).  Its ``middle" fraction $r_M=r_{k+2}^{(4n-1)}$ is of odd order, and was therefore carried over from level $k+1$, where it is indexed as $r_M=r_{k+1}^{(2n)}$.  This fraction is even, and therefore newborn at level $k+1$.  Hence the neighboring fractions to its left and right, $r_{k+1}^{(2n-1)}$ and $r_{k+1}^{(2n+1)}$, respectively, are odd.  These two fractions, therefore, appear at level $k$ as $r_{k}^{(n)}$ and $r_{k}^{(n+1)}$, respectively.

Now the denominators of odd order fractions carry over from the previous level, i.e., $d_{k+1}^{(2n-1)}=d_{k}^{(n)}$, while those at even order (since they belong to ``new" Farey fractions) are the sum of their neighbors, i.e., $d_{k+1}^{(2n)}=d_{k+1}^{(2n-1)}+d_{k+1}^{(2n+1)}=d_{k}^{(n)}+d_{k}^{(n+1)}$.
\begin{figure}[h]
\centering\includegraphics[width=10.0cm]{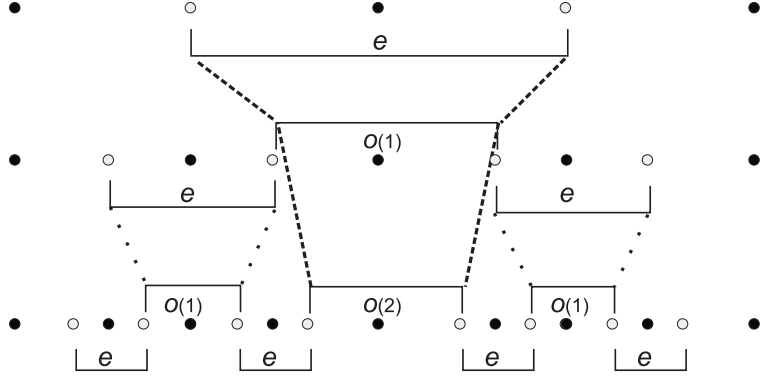}
\caption{Even and odd interval lineage.  The numbers indicate age of a given odd interval.}\label{LMR}
\end{figure}

Putting these things together with  Lemma \ref{lem3} gives
\begin{lem} \label{lem4}
For any $k \ge 0$ and  $n=1, 2, \ldots, 2^k$, i.e., any even interval at level $k+2$, we have
\begin{eqnarray}\label{LengthEI}
I_{k+2}^{(2n)}=\frac{3}{(d_{k}^{(n)}+2d_{k}^{(n+1)})(2d_{k}^{(n)}+d_{k}^{(n+1)})}\;,
\end{eqnarray}
furthermore, for any
 $l \ge 1$,
\begin{eqnarray}\label{Lengthl}
\;I_{k+l+1}^{(2^{l-1}(2n-1)+1)}=\frac{2l+1}{(ld_{k}^{(n)}+(l+1)d_{k}^{(n+1)})((l+1)d_{k}^{(n)}+ld_{k}^{(n+1)})}\;,
\end{eqnarray}
where, for $l > 1$, (\ref{Lengthl}) includes, at level $k+l+1$, all descendants of the $2^k$ even intervals at level $k$.
\end{lem}
 Lemma \ref{lem4} is illustrated in  Fig. \ref{LMR}.

Now (except for $I_k^{(1)}$) every odd interval (for $k > 2$) is the descendant of a unique even interval at some lower level.  Therefore (\ref{Lengthl}) is valid for all $2^{k+l-1}-1$ odd intervals at any level $k+l+1$ with $k \ge 0$ and $ l > 1$, omitting $I_k^{(1)}$.  Consider the identity $2^{k+l-1}-1 = \sum_{i=2}^{k+l}2^{i-2}$.  In this context it expresses the number of odd intervals at level $k+l+1$ in terms of a sum over the numbers of (even) ``parent" intervals at each lower level $i$, with $2 \le i \le k+l$.

At this point,  we consider the ratio of an arbitrary odd interval, as given by (\ref{Lengthl}), to its parent interval  $I_{k+2}^{(2n)}$.  For simplicity, we let $m=k+2$  and $j=l-1$, so that $m \ge 2$ and $j > 0$, and relabel the lhs of (\ref{Lengthl}) as $I^{([2n,j])}_{m+j}$, to indicate that it is the $j$th descendant of the $2n$th interval at level $m$ (note that $I^{([2n,0])}_{m} = I^{(2n)}_{m}$).  Then, with $z := \frac{d_{k}^{(n)}}{d_{k}^{(n+1)}}$ we find 
\begin{lem} \label{lem5}
For $m \ge 2$ and $j>0$,
\begin{eqnarray}\label{ratioOfInt}
\frac{I^{([2n,j])}_{m+j}}{I_{m}^{(2n)}}=\frac{(2j+3)}3 \frac{(1+2z)(2+z)}{((j+1)+(j+2)z)((j+2)+(j+1)z)}\;.
\end{eqnarray}
\end{lem}
 Lemma \ref{lem5} expresses each successive descendent odd interval in terms of its parent even interval.  It leads immediately, via elementary computations, to

\begin{lem} \label{lem6}
For $j>0$ and $m \ge 2$,
\begin{eqnarray}\label{ratiobds}
\frac{2(2j+3)}{3(j+1)(j+2)} \le \frac{I^{([2n,j])}_{m+j}}{I_{m}^{(2n)}} \le \frac3{2j+3}.
\end{eqnarray}  
\end{lem} \qed \\ 

The lower bound in (\ref{ratiobds}) arises from (\ref{ratioOfInt}) at $z=0$ or $z=\infty$, and  the upper bound  from  (\ref{ratioOfInt}) at $z=1$.  In the case of Farey fraction denominators none of these $z$ values actually 
occurs. However, the important point is that these bounds are  independent of both the parent level $m$ and the initial (parent) even interval. 

Let $I^{([o,j])}_{m+j}$ denote the sum of all odd intervals at level $m+j$ that are descendants of the even intervals at an arbitrary level $m \ge 2$. Then
\begin{lem} \label{lem7}
For $j>0$ and $m \ge 2$ 
\begin{eqnarray}\label{SumRatio}
I^{(e)}_m\frac{2}{3}\frac{(2j+3)}{(j+1)(j+2)} \le I^{([o,j])}_{m+j} \le I^{(e)}_m \frac3{2j+3}\;.
\end{eqnarray}  
\end{lem}
{\bf Proof:}
Sum (\ref{ratiobds}) over the even intervals at level $m$, and make use of the definition (\ref{EvenInt}). \qed
\\ \\
{\bf Proof of Lemma \ref{lem2}:}  The remainder of the proof is as follows.  First, relabel the ``parent" level in   Lemma \ref{lem7} as $m = k - j$. If we fix $k=m+j > 2$, since $m$ varies over the range $2 \le m \le k-1$, $j$ satisfies $1 \le j \le k-2$.  Thus all the odd intervals at an arbitrary level $k$ are included, except the ``end" interval $I^{(1)}_k$.  The leftmost inequality in (\ref{OddIntB}) and the inequality in (\ref{OddIntUpB}) then follow directly on summing (\ref{SumRatio}) and using  (\ref{Int1}). \qed

Finally,   A. Zhigljavsky \cite{Z} has verified numerically, up to level $k=34$, that  $I_k^{(e)}$ continues to decrease as  $k$ increases.  His results are consistent with the result that $I_k^{(e)} \sim 1/\log_2(k)$ as $k \to \infty$ found in \cite{KS}.

\section{Discussion}\label{disc}
\subsection{Proving the Conjecture} \label{PtC}
The conjecture (\ref{limI}) is, as already pointed out, very simple.  However a proof is apparently quite elusive, at least using the methods employed in this paper. Even establishing that $I^{(e)}_k$ is monotonically decreasing with $k$, which, given (\ref{liminfI}), would be sufficient, appears very non-trivial. However several recent approaches to proving  this conjecture have been proposed (\cite{Z,M,KS}), based, respectively, on the Chacon-Ornstein ergodic theorem, continued fractions, and a measure theoretic analysis.  

\subsection{Relation to Physics} \label{relP}

The problem treated here arose from previous investigations of the Farey fraction spin chains, a set of statistical mechanical models based on the Farey fractions (see \cite{KO}, \cite{FKO} and references therein for details).    It follows directly from their definitions  that the ``Farey tree partition function" $Z_{k}^F(\beta)$ (see  equation (7) in \cite{FKO}) satisfies $Z_{k}^F(1) = I_k^{(e)}$, while the ``even Knauf partition function" $Z_{k,e}^K(\beta)$ (see the equation after (3) in \cite{FKO}) satisfies $Z_{k,e}^K(2) = S_k$.  

Therefore the inequality (13), proven  in \cite{FKO}, can be rewritten as 
 \begin{equation}\label{bds}
S_k < I_k^{(e)} < 4 S_{k-1} \; ,
\end{equation}
which immediately proves  
\begin{lem} \label{lem1}
 The conjecture (\ref{limI}) is equivalent to
\begin{equation}  \label{limS}
\lim_{k \to \infty} S_k=0 \;.  
\end{equation}
\end{lem} \qed

Note that the term ``partition function" is used in its statistical mechanical sense here, which in general has no connection with the number theoretic usage.

Since this paper was written, as mentioned in Section \ref{PtC}, several proofs of Conjecture (\ref{limI}) using different methods have been proposed.    The method employed here does not seem capable of establishing  Conjecture (\ref{limI}), however it gives detailed information about the evolution of the intervals not otherwise available.  

There are several spin chains known to have the same free energy, and thus the same thermodynamic behavior.  They all  exhibit a second-order phase transition at non-zero temperature. (The free energy  is defined via $f(\beta):=\lim_{k \to \infty} \left( -\log Z_{k}^F(\beta)/k \beta\right)$, and a phase transition is a singularity in $f(\beta)$ -- see \cite{FKO} for   more details on these matters.)  The Farey tree model is employed in \cite{FPT} for a study of multifractal behavior associated with chaotic maps exhibiting intermittency.   The critical point (phase transition) in this model  occurs at $\beta=1$.  Therefore, physically, $Z_k^F(1)=I^{(e)}_k$ is the value of the partition function at the critical point.  (The value of the partition function at one point is, however, generally of no physical interest.)

 In proving  that the free energy of the Farey tree model is the same as the free energy for other Farey statistical models, it was already demonstrated \cite{FKO} that as $k \to \infty$, $Z_k^F(\beta)\to 0$ for $\beta > 1$, while $Z_k^F(\beta)\to \infty$ for $\beta < 1$. (This, incidentally, also establishes that the Hausdorff dimension is $\beta_H = 1$.)  However, exactly at the critical point, it was only shown that $0 < Z_k^F(1) <1$. To our knowledge, the result (\ref{liminfI}) found here, and extended by \cite{Z,M,KS}, is new.  
 
 Finally, we note that \cite{MZ} contains some related work, giving results on the large $k$ behavior of the quantity \begin{equation}
\sigma_k(\beta):= \sum_{i=1}^{2^k}\left( r_k^{(i+1)}- r_k^{(i)} \right)^{\beta} \;,
\end{equation} 
for $\beta >1$.

\section*{Acknowledgements}\label{fp}
 
We are grateful to Don Zagier and A. Zhigljavsky for conversations, and to the latter for numerical results. 

 The second author was supported in part by the National Science Foundation under Grants Nos. DMR-0203589, DMR-0536927 and  PHY99-07949. He is also grateful for the hospitality of the Kavli Institute for Theoretical Physics, University of California, Santa Barbara, where part of this work was done.

\begin{otherlanguage}{english}

\end{otherlanguage}

\end{document}